\documentclass[structabstract]{aa}  
\usepackage{graphicx}
\usepackage[varg]{txfonts}
\usepackage{longtable,lscape}
\usepackage[]{natbib}
\bibpunct{(}{)}{;}{a}{}{,}

\begin{document}
   \title{Pre-ALMA observations of GRBs in the mm/submm range
   \thanks{This publication is partially based on data acquired with the Atacama Pathfinder Experiment (APEX) under programmes 082.F-9850, 084.D-0732, 086.D-0590, 086.F-9303(A) and 087.F-9301(A) and with the Submillimeter Array (SMA) under programmes 2009B-S015, 2010A-S004 and 2010B-S026. This paper makes use of the following ALMA commissioning data set: 2011.0.99001.CSV.}
   }
   \author{
   A.~de~Ugarte~Postigo \inst{1,2}
          \and
   A.~Lundgren \inst{3,4}
          \and
   S.~Mart\'in \inst{3}
          \and
   D.~Garcia-Appadoo \inst{3,4}
          \and
   I.~de~Gregorio Monsalvo\inst{3,4}
	\and
   A.~Peck\inst{4}   
   	\and
   M.~J.~Micha{\l}owski \inst{5}
 	\and
   C.~C.~Th\"one\inst{2}
 	\and
   S.~Campana\inst{6}
 	\and
   J.~Gorosabel\inst{2}
 	\and
   N.~R.~Tanvir\inst{7}
 	\and
   K.~Wiersema\inst{7}
 	\and
   A.~J.~Castro-Tirado\inst{2}
 	\and
   S.~Schulze\inst{8}
     	\and
   C.~De~Breuck\inst{9}
   	\and
   G.~Petitpas\inst{10}
         \and
   J.~Hjorth\inst{1}
	\and
   P.~Jakobsson\inst{8}
 	\and
   S.~Covino\inst{6}
 	\and
   J.~P.~U.~Fynbo\inst{1}
   	\and
   J.~M.~Winters\inst{11}
   	\and
   M.~Bremer\inst{11}
         \and
   A.~J.~Levan\inst{12}
         \and
   A.~Llorente\inst{13}
   	\and
   R.~S\'anchez-Ram\'irez\inst{2}
	\and
   J.~C.~Tello\inst{2}
	\and
   R.~Salvaterra\inst{14}
          }

   \institute{
   	Dark Cosmology Centre, Niels Bohr Institute, Juliane Maries Vej 30, Copenhagen \O, 2100, Denmark\\
              \email{adeugartepostigo@gmail.com}
         \and
	Instituto de Astrof\'isica de Andaluc\'ia (IAA-CSIC), Glorieta de la Astronom\'ia s/n, 18008, Granada, Spain 
	\and
   	European Southern Observatory, Vitacura Casilla 19001, Santiago de Chile 19, Chile
	\and
	Joint ALMA Observatory, Alonso de C\'ordova 3107, Vitacura - Santiago, Chile
	\and
	Scottish Universities Physics Alliance, Institute for Astronomy, University of Edinburgh, Royal Observatory, Edinburgh, EH9 3HJ UK
 	\and
	INAF - Osservatorio Astronomico di Brera, via E. Bianchi 46, 23807 Merate (LC), Italy
 	\and
	Department of Physics and Astronomy, University of Leicester, University Road, Leicester LE1 7RH, UK
	\and
             Centre for Astrophysics and Cosmology, Science Institute, University of Iceland, Dunhagi 5, IS-107 Reykjav\'ik, Iceland
	\and
	European Southern Observatory, Karl Schwarzschild Stra§e 2, 85748, Garching, Germany
	\and
	Harvard-Smithsonian Center for Astrophysics, Submillimeter Array, 645 North A'ohoku Place, Hilo, HI 96720, USA
	\and
	Institut de Radioastronomie Millim\'etrique (IRAM), 300 rue de la Piscine, F-38406 Saint Martin d'H\`eres, France
	\and
	Department of Physics, University of Warwick, Coventry, CV4 7AL, UK
	\and
	Herschel Science Operations Centre, INSA, ESAC, Villafranca del Castillo, 50727, 28080 Madrid, Spain
	\and
	INAF/IASF Milano, via E. Bassini 15, I-20133 Milano, Italy
             }

   \date{Received; accepted }

  \abstract
   {
   Gamma-ray bursts (GRBs) generate an afterglow emission that can be detected from radio to X-rays during days, or even weeks after the initial explosion. The peak of this emission crosses the millimeter and submillimeter range during the first hours to days, making their study in this range crucial for constraining the models. Observations have been limited until now due to the low sensitivity of the observatories in this range. This situation will be greatly improved with the start of scientific operations of the Atacama Large Millimeter/submillimeter Array (ALMA).}
   {In this work we do a statistical analysis of the complete sample of mm/submm observations of GRB afterglows obtained before the beginning of scientific operations at ALMA.}
   {We present observations of 11 GRB afterglows obtained from the Atacama Pathfinder Experiment (APEX) and the SubMillimeter Array (SMA), as well as the first detection of a GRB with ALMA, still in the commissioning phase, and put them into context with a catalogue of all the observations that have been published until now in the spectral range that is covered by ALMA.}
   {The catalogue of mm/submm observations collected here is the largest to date and is composed of 102 GRBs, of which 88 have afterglow observations, whereas the rest are host galaxy searches. With our programmes, we contributed with data of 11 GRBs and the discovery of 2 submm counterparts. In total, the full sample, including data from the literature, has 22 afterglow detections with redshifts ranging from 0.168 to 8.2.
   GRBs have been detected in mm/submm wavelengths with peak luminosities spanning 2.5 orders of magnitude, the most luminous reaching $10^{33} \rm{erg}$ $\rm{s}^{-1} \rm{Hz}^{-1}$.
   We observe a correlation between the X-ray brightness at 0.5 days and the mm/submm peak brightness. Finally we give a rough estimate of the distribution of peak flux densities of GRB afterglows, based on the current mm/submm sample.
   }
   {Observations in the mm/submm bands have been shown to be crucial for our understanding of the physics of GRBs, but have until now been limited by the sensitivity of the observatories. With the start of the operations at ALMA, the sensitivity has improved by more than an order of magnitude, opening a new era in the study of GRB afterglows and their host galaxies. Our estimates predict that, once completed, ALMA will detect up to $\sim$ 98\% of the afterglows if observed during the passage of the peak synchrotron emission.    
   }


   \maketitle
%

\section{Introduction}

Gamma-ray bursts (GRBs) are the brightest explosions in the Universe,
releasing isotropic equivalent energies in the range of $10^{51}$ -- $10^{54}$ ergs 
(of the order of $10^{51}$ ergs once corrected for collimation) within a few seconds.
They were serendipitously discovered in 1967 \citep{kle73} in gamma-rays, but it was not
until 1997 that the first counterparts were detected at other wavelengths \citep{van97,fra97,cos97,bre98}.
Thanks to the observation of the fading X-ray emission, which follows the 
more energetic gamma-ray photons, and the rapid distribution of the X-ray coordinates
(much more precise than the gamma-ray ones), it is possible to carry out 
multiwavelength observations of the counterparts associated with GRBs within 
seconds of their occurrence. The late time emission, detectable at all wavelengths and observable 
for days after the burst onset (and some times significantly longer),
is what we call GRB \textit{afterglow}.

It is widely accepted that long GRBs (those with durations above $\sim$2 s) 
are related to the collapse of massive stars \citep{woo93,pac98,hjo03} while
short GRBs (shorter than $\sim$2 s) are probably originated by the coalescence
of two compact objects, such as two neutron stars or a neutron star and 
a black hole \citep{pac90,nar92,lev06,cha07,kin07}. In both scenarios, the result is the ejection of material
at ultrarelativistic speeds, most probably through jets. This is what we 
call the \textit{relativistic fireball model} \citep{ree92,sar98}. In this framework, a compact source releases $\sim$ 10$^{51}$ 
ergs of energy within dozens of seconds in a region smaller than 10 km. 
When the ejecta run into the surrounding medium, a forward shock
sweeps the surroundings of the progenitor, producing an afterglow as the ejecta 
interact with the interstellar matter. A reverse 
shock, colliding with the ejecta, can also produce an additional emission during
early times \citep{pir99}.

Until now, millimeter and especially submillimeter observations of GRBs have been scarce due to the small
number of available observatories and limited sensitivity, as compared to other wavelength ranges. As examples of large samples of afterglow data see, for example, \citet{eva09} in X-rays, \citet{kan10} in optical, and more recently \citet{cha11} in radio.
The mm/submm range suffers strongly from atmospheric absorption, mostly due to water vapour. This means that
observatories have to be placed in high and dry mountain sites. Nevertheless, the study of GRBs in these wavelengths 
is of great interest, as the peak (in flux density) of the synchrotron afterglow emission, and even the one due to the 
reverse shock (see Sect.~\ref{sect:fb} for a description of the emission mechanisms), is expected to be located in this wavelength range
during the first days. It has also the benefit of being a wavelength range that is not affected by interstellar extinction, as are the
optical or soft X-rays, allowing us to detect highly-extinguished and high-redshift bursts. This range is also normally above the 
self-absorption frequency of the spectrum, below which the flux is strongly suppressed. Finally, as compared to centimeter 
and decimeter wavelengths, this range is not affected by interstellar scintillation, allowing more accurate modelling with fewer observations.

Pioneering submm observations of GRB afterglows, made with
JCMT/SCUBA \citep[e.g.][]{smi99,smi01} in the submm
and with various facilities in the mm \citep[e.g.][]{bre98,gal00},
showed that emission in these wavelengths could be detected in some cases, and
provided the first tests of the fireball model using full radio to X-ray SEDs
\citep[e.g.][]{wij99}.
However, detections were generally of low significance, and progress in the
\emph{Swift} era was restricted until recently by the dearth of submm bolometer array
instruments (following the retirement of SCUBA on JCMT, and until the advent of
LABOCA on APEX).

On the 30$^{\rm{th}}$ September 2011 ALMA came into operation, to revolutionise mm/submm
astronomy. ALMA is at an altitude of 5000 m in the Llano de Chajnantor, in northern Chile, one
of the best sites for this kind of observations in the world. In its early science phase it has 16 antennas of 12 m, and in full operations (expected for 2013) it will increase to
a total of 50 antennas of 12 m in diameter plus a compact array (the Atacama Compact Array - ACA) of 12 7 m antennas and four 12 m antennas.
This will be the largest facility for mm/submm astronomy and will improve sensitivity and spatial resolution in over an order of 
magnitude with respect to previous observatories.

In this paper we collect all the data, to our best knowledge, that have been published for GRBs in the mm/submm range, 
to which we add observations from our observing programmes from APEX and the SMA. The observations are completed with the first ALMA observation of a GRB, during the commissioning phase of the observatory, as a test of the target of opportunity procedures. Using this sample, we give a review of the 
past experience and put it into context of the forthcoming ALMA era. We give some examples of single-epoch 
spectral energy distributions but do not attempt full broadband modelling for individual bursts as it is beyond the scope of this paper.

Section 2 describes the data of 11 bursts collected with our observing programmes. Section 3 presents the catalogue of mm/submm GRB data.
In Section 4 we describe the capabilities of ALMA compared with previous observatories and discuss its future role in the field. Finally, Section 5 lists our conclusions. Throughout the paper we assume a cosmology with $\Omega_m=0.30$, $\Omega_\Lambda=0.70$, and $H_0=70$.


   \begin{figure}[h]
   \centering
   \includegraphics[width=8cm]{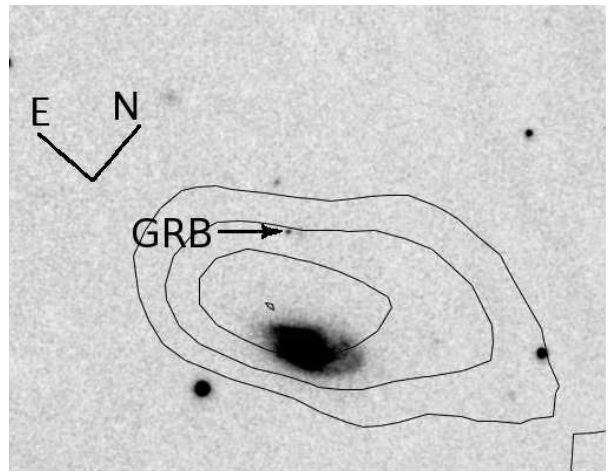}
      \caption{Observation of GRB\,090313 with APEX (contours are 1, 2, 3 and 4$\sigma$ detection levels) in March 2009 plotted over an optical X-shooter acquisition image showing the position of the afterglow. The field of view is $90^{\prime\prime}\times70^{\prime\prime}$. Only a field galaxy, unassociated with the GRB is detected, and we impose a 3-sigma upper limit of 9 mJy for the afterglow \citep[adapted from ][]{deu10e,mel10}.
              }
         \label{Fig:090313}
   \end{figure}

\section{Observations}

     \begin{figure*}
   \centering
   \includegraphics[width=18cm]{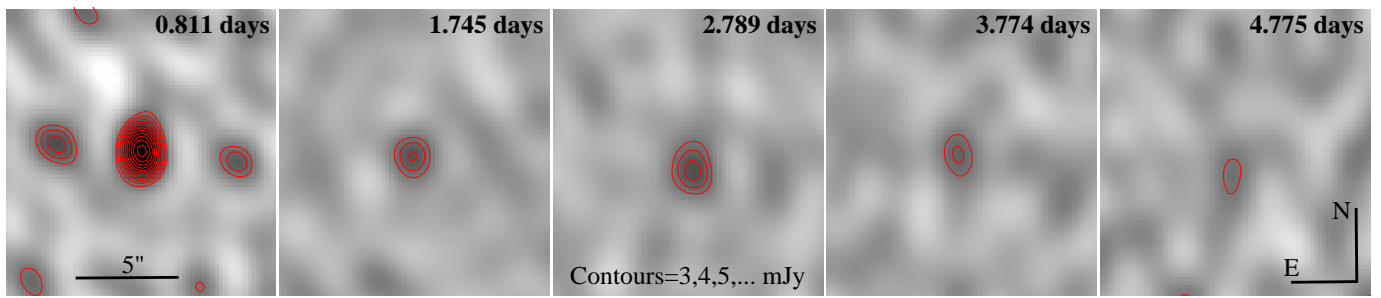}
      \caption{Observations of GRB\,100418A obtained with SMA during the first 5 days after the GRB \citep[adapted from][]{deu11b}.
              }
         \label{Fig:100418a}
   \end{figure*}

On March 2009 we performed the first observation of a GRB (GRB\,090313) using the 12 m, single-dish, APEX telescope \citep[in Chajnantor, Chile;][]{gus06} at 345 GHz, as a feasibility test, during a technical night through a Director Discretionary Time programme. Since then, we have had several observing programmes at APEX, with which we have followed-up 6 GRBs (plus an additional Galactic X-ray binary, initially identified as a GRB). All these observations were performed using the Large Apex BOlometer CAmera \citep[LABOCA;][]{sir09}, a multi-channel bolometer array with 295 elements designed for continuum observations, with a field of view (FOV) of $11\farcm4\times11\farcm4$ and an angular resolution of $19.5\pm1^{\prime\prime}$. Observations until the end of 2009 were performed using a spiral raster mapping, whereas later observations used the photometric mode. The mapping mode results in a fully sampled and homogeneously covered map which is less sensitive than the photometric mode (which lacks the spatial resolution) that was made available in 2010. We chose to use the photometric mode when it became available, as we are more interested in the sensitivity than in the spatial coverage, as our sources are unresolved. Data reduction of LABOCA/APEX observations was done using BoA\footnote{http://www.mpifr-bonn.mpg.de/div/submmtech/software/boa/\\boa\_main.html}, as well as CRUSH and miniCRUSH \citep{kov08} software packages.

Since November 2009 we have an observing project for GRB afterglows from the northern hemisphere at the 8-antenna SMA interferometer (Hawaii, U.S.A.) with which we have followed-up 4 other bursts. Observations from SMA were reduced using the MIR-IDL\footnote{https://www.cfa.harvard.edu/$\sim$cqi/mircook.html} and MIRIAD \citep{sau95} packages.

In the following, we provide details on each of our GRB observations. The data are listed, together with the complete catalogue in Table~\ref{table:mmdata}.

\textbf{GRB\,090313:} This burst was a 78 s long event with a bright afterglow at a redshift of $z=3.3736$ \citep{deu10e}. After the discovery of a bright mm counterpart \citep{boc09}, we obtained continuum observations at 345 GHz using LABOCA/APEX bolometer array during technical time. Data were acquired on 2009 March 17 and 24 under good weather conditions (zenith opacity values ranged from 0.24 to 0.33 at 345 GHz). Observations were performed using a spiral raster mapping, providing a fully sampled and homogeneously covered map in an area of diameter $12^{\prime}$, centered at the coordinates of the optical afterglow of GRB\,090313. The total on-source integration time of the two combined epochs was 4.6 hr. Calibration was performed using observations of Saturn as well as IRAS 09452+1330, HD 82385, G10.6-0.4, and IRAS 17574-2403 as secondary calibrators. The absolute flux calibration uncertainty is estimated to be 11\%. The telescope pointing was checked every hour, finding a root mean square (r.m.s.) pointing accuracy of $1\farcs8$. The individual maps were co-added and smoothed to a final angular resolution of $27\farcs6$. We obtained a 3$\sigma$ upper limit of 14 mJy for each of the two epochs and a limit of 9 mJy in the coadded maps (see Fig.~\ref{Fig:090313}). These data were presented, together with an analysis of the evolution of the afterglow emission by \citet{mel10}.

\textbf{GRB\,091102:} This was the first burst that we observed during our regular programme time. It was a 6.6 s long burst, for which no credible optical counterpart was identified and no redshift was obtained. LABOCA/APEX observations were performed on 2009 November 3 under average weather conditions (zenith opacity values ranged from 0.4 to 0.5 at 345 GHz). The total on-source integration time was $\sim$2 hr. Calibration was performed using observations of Uranus and Neptune. The absolute flux calibration uncertainty is estimated to be $\sim$ 8$\%$. The telescope pointing was checked every hour towards the source PMNJ0450-8100. We obtained a 3$\sigma$ point source sensitivity of 20 mJy.

\textbf{GRB\,091127:} This was a nearby burst \citep[$z=0.49044$;][]{ver11} with a duration of 7.1 s and a bright optical counterpart that we followed with LABOCA/APEX on 2009 Nov 28 and 29 at 345 GHz. Weather conditions were average, with zenith opacity values ranging from 0.4 to 0.63 at 345 GHz. Observations were performed using a spiral raster mapping. The total on-source integration time of the two combined epochs was 6.4 hr (2.9 and 3.5 hr in the first and second epoch, respectively). Pointing was checked regularly on J0050-095 and J0145-276. Calibration was performed using observations of Uranus and the secondary calibrator NGC 2071 IR. The absolute flux calibration uncertainty is estimated to be about 15\%. We obtained 3$\sigma$ upper limits of 14.9 mJy and 13.2 mJy for each of the two epochs, respectively, and 9.6 mJy for the combined epochs. An analysis of the afterglow emission of this burst, including the APEX data, is presented by \citet{ver11}.

\textbf{GRB\,100418A:} On April, 18th 2010, the \emph{Swift} Burst Alert Telescope (BAT) triggered and located GRB 100418A. It was an intermediate duration burst (7.0 s) at a redshift of $z=0.62$ \citep{ant10}. It had a peculiar optical light curve peaking several hours after the event. We performed observations of the afterglow using 7 out of the 8 SMA antennas, starting $\sim$ 16 hr after the burst.
Weather conditions were good, with zenith opacities at 225~GHz of $\tau\sim0.06$ (precipitable water vapour, PWV~$\sim$1~mm).
Titan and Neptune were used as flux calibrators and 3C454.3 as bandpass calibrator. Atmospheric gain was corrected from observations of the 
nearby quasar J1751+096 every 15 min. Using these data we discovered a bright counterpart \citep{mar10} at a flux of 13.40$\pm$1.60 mJy (see Fig.~\ref{Fig:100418a}). At the time of the discovery, this was the second brightest mm/submm counterpart detected (after GRB\,030329, which peaked at $\sim$70 mJy; \citealt{she03,res05}). Observations continued over the following 4 nights, tracking the evolution of the afterglow until it became undetectable on April 23rd \citep{deu11b}.

\textbf{GRB\,100814A:} This was a very long burst (150 s) with a bright counterpart at a redshift of $z=1.44$ \citep{ome10}. Continuum observations at 345 GHz were carried out using LABOCA/APEX. Data were acquired on August 15 2010, starting 26 hr after the burst, under good weather conditions (zenith opacity value was 0.28 at 345 GHz). Observations were performed using the photometry mode. The total on-source integration time was 1.5 hr. Pointing was checked regularly on J0145-276. Calibration was performed using observations of Uranus and the secondary calibrator V883-ORI. The absolute flux calibration uncertainty is estimated to be about 15\%. The formal flux measured at the position of the afterglow was -0.33$\pm$1.6 mJy/beam, i.e. a 3$\sigma$ limit of 4.8 mJy. A radio counterpart was detected by \citet{cha10} several days later from EVLA.

\textbf{GRB\,100901A:} This was a bright event, at a redshift of $z$=1.41 \citep{cho10}. Observations were carried out with SMA on the 3rd of September 2010. All eight antennas, arranged in the extended array configuration, were tuned to 345.8~GHz. Weather was excellent with a zenith opacity at 225~GHz of $\tau\sim 0.06$ (precipitable water vapour, PWV~$\sim 1$~mm). Uranus and Callisto were observed as flux calibrators and 3C454.3 as bandpass. Atmospheric gain calibration was derived from the observations of the nearby quasars J0237+288 and J0319+415 every 15 min. The GRB was observed for $\sim 10$ hr, and the observations resulted in a non detection, being the GRB flux constrained down to an r.m.s. noise of $\sim 0.75$ mJy, which gives the deepest limit in our sample. A radio counterpart was later detected by WSRT \citep{van10b,van10c} and EVLA \citep{cah10b}.

\textbf{GRB\,100925A/MAXI J1659-152:} On 2010 September 25 we responded to a GRB alert using APEX. Observations were performed 15.5 hr after the burst onset and a bright submillimeter afterglow was detected, with a flux density of 12.6$\pm$2.4 mJy \citep{deu10c}. Later spectroscopic observations performed with the X-shooter spectrograph at the Very Large Telescope, showed that the source was not due to an extragalactic GRB but to an unusually energetic burst from an X-ray binary \citep{deu10d}. Thanks to the detection from APEX, an extensive radio observation campaign was triggered \citep[including further observations from APEX]{van10}.

\textbf{GRB\,110422A:} Observations of this burst at $z=1.770$ \citep{mal11,deu11e} were carried out with the SMA on May 6th, 2011 in search for an emission tentatively detected by the \emph{Herschel Space Observatory} \citep{hua11}. The seven available antennas were tuned to 233.7~GHz.
Weather was bad but stable enough with zenith opacity at 225~GHz was $\tau\sim0.35$ (PWV~$>5$~mm). J0721+713 was used as flux calibrator and 3C273 as bandpass. Atmospheric gain calibration was derived from J0721+713 quasar observations every 15 min. The GRB was observed for 4 hr, reaching an r.m.s. noise of $\sim2.8$ mJy but obtained no detection.

\textbf{GRB\,110503A:} Observations of this $z=1.613$ \citep{deu11f} burst were carried out with the SMA on May 4th, 2011. The five available antennas were tuned to 225.0~GHz.
Weather was mediocre but stable with zenith opacity at 225~GHz was $\tau\sim0.25$ (PWV~$>4$~mm).
J0927+390 was used as flux calibrator and 3C273 as bandpass. Atmospheric gain calibration was derived from J0927+390 and
J0920+446 quasar observations every 15 min. The GRB was observed for 7 hr reaching an r.m.s. noise of $\sim1.7$ mJy.

\textbf{GRB\,110709B:} This was a very peculiar burst, that showed two gamma-ray triggers over a period of almost 15 minutes \citep{bar11}. Early optical observations showed that it was a very dark event \citep{fon11}, with no credible afterglow detected or redshift measured. We performed submm observations with APEX to try to localise the counterpart and to constrain the synchrotron spectrum. Observations began on July 11, 2.00 days after the burst, with an on-source time of 106 min and precipitable water vapour of 0.8 mm. We did not detect any significant emission, obtaining a formal flux on the source position of 1.4$\pm$2.3 mJy (3-sigma limit of 6.9 mJy). A detection of a radio counterpart at 5.8 GHz was later reported by \citet{zau11d}.

\textbf{GRB\,110715A:} This burst had a very bright optical counterpart in spite of being located very close to the plane of the Milky Way from which it suffered from significant dust extinction. Its redshift was measured to be $z=0.82$ \citep{pir11}. APEX observations began on July 16, 1.42 days after the burst and were performed using the photometric mode. The weather conditions were very good, with a precipitable water vapor of 0.62 mm. Using these observations we discovered a bright submm counterpart at 10.4$\pm$2.4 mJy \citep{deu11d}.


As a test of the target of opportunity procedure, GRB\,110715A was subsequently observed at Atacama Large Millimeter Array (ALMA). The ALMA Science Team report a preliminary detection from a test observation of this source of 4.9$\pm$0.6 mJy at 345 GHz after 25 mins on source with 7 antennas.
 The centroid of the ALMA position is 15:50:44.05 -46:14:06.54 with an uncertainty of $0\farcs3\times0\farcs1$ at a position angle of 76 degrees. Observations began on July 19 at 02:50 UT (3.57 days after the burst). The weather conditions were very good, with a precipitable water vapor of 0.5 mm. In spite of being obtained during a test observation, with an order of magnitude fewer antennas than will be available with the full observatory and for only 25 minutes, this is the deepest observation in the complete sample at 345 GHz and provides the most accurate coordinates available for this burst. A complete analysis of the afterglow emission of this burst, will be presented by \citet{san11}.

\textbf{GRB\,110918A:} This burst, at a redshift of $z=0.98$ \citep{lev11} was one of the brightest GRBs ever detected in gamma-rays, and the brightest ever observed by Konus/\textit{WIND} \citep{gol11}. The large error box generated just with Konus/\textit{WIND} data did not allow us to observe until 2 days after, when the coordinates were refined thanks to the detection of X-ray \citep{man11} and optical counterparts \citep{tan11}. Continuum observations at 345 GHz were carried out using LABOCA/APEX. Data were acquired on 2011-09-21 between 03:02 and 05:46 under excellent conditions (zenith opacity value was 0.2 at 345 GHz). Observations were performed using the photometry mode. The total on source integration time was  2.41 hours. Pointing was checked regularly on J0145-276. Calibration was performed using observations of Uranus and Neptune and and the secondary calibrator V883-ORI. The absolute flux calibration uncertainty is estimated to be about 15\%. Because of technical issues, the data did not reach the theoretical noise level, being no flux measured at the position of the afterglow and the upper 3-$\sigma$ limit was 15 mJy. 

\section{The pre-ALMA millimeter/submillimeter sample of GRBs}
\label{sec:pre}

   \begin{figure}
   \centering
   \includegraphics[width=9cm]{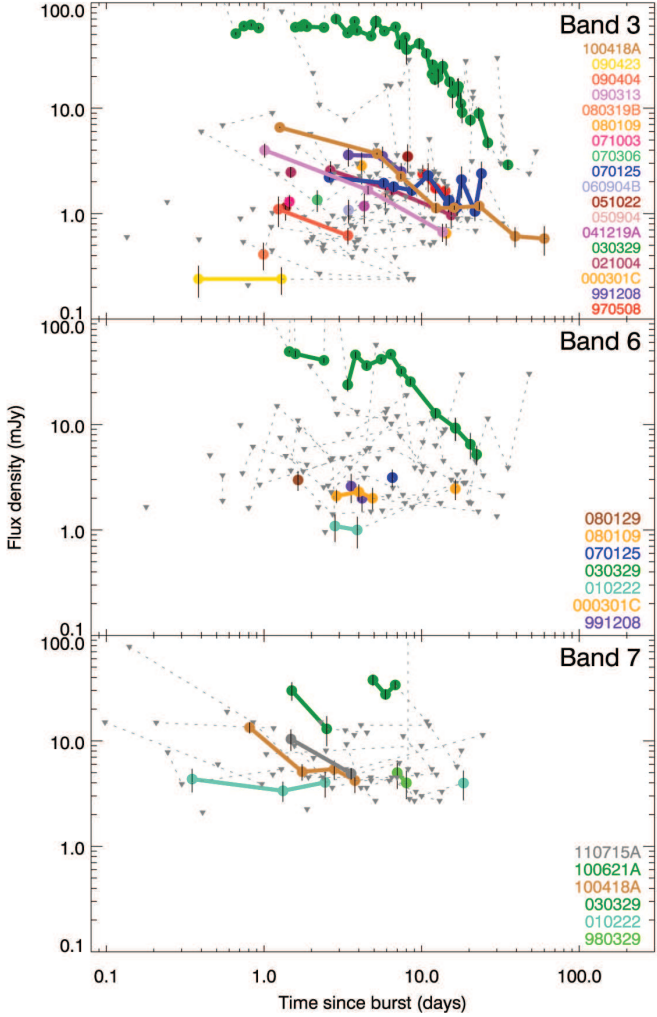}
      \caption{Light curves of GRB afterglows in the different bands (see Table~\ref{table:bands} for a definition of the bands). Coloured dots indicate detections, while gray triangles are 3$\sigma$ detection limits. Observations of an individual burst are connected by a thick coloured line in case of detections and a thin dotted line in case of detection limits. GRBs with only one observation are shown as individual symbols with no connecting lines. Detections of GRB\,010222 in bands 6 and 7 are due to the host galaxy and not the afterglow. The detection of XT\,080109 in band 6 is due to the host galaxy.
              }
         \label{lcs}
   \end{figure}
   
      \begin{figure}
   \centering
   \includegraphics[width=9cm]{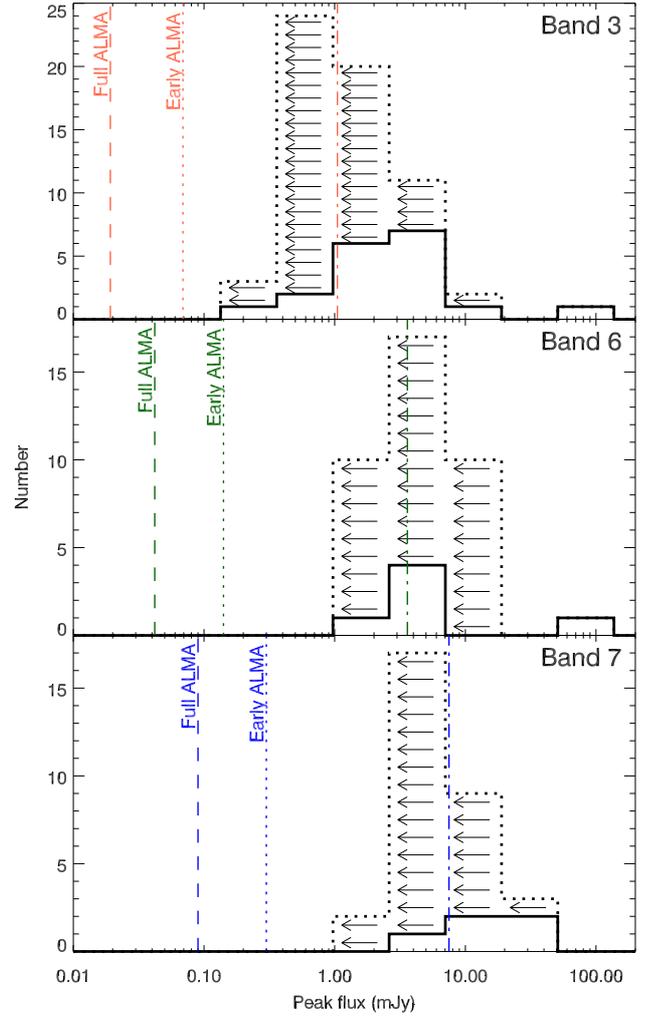}
      \caption{Histograms of peak fluxes and 3$\sigma$ upper limits in each of the three main bands in our sample. Detection limits for early and full ALMA (3$\sigma$, derived from the r.m.s. calculated in Table~\ref{table:rms}) are indicated with vertical dotted and dashed lines, respectively. The dashed-dotted line indicates the median detection limit of the early observations (see Table~\ref{table:1st}).
              }
         \label{hist}
   \end{figure}

In order to put our observations into context, we have collected, in Table~\ref{table:mmdata}, the most complete sample of continuum observations that have been published to date of GRB afterglows and their host galaxies in the ALMA wavelength range, covering from early 1997 until the 30$^{th}$ of September 2011 (starting day of the early science operations with ALMA). The data are ordered chronologically by GRB, considering as detections only those with higher significance than 3$\sigma$, and otherwise providing detection limits. The observations are separated into ALMA bands 3 (84 -- 116 GHz), 4 (125 -- 163 GHz), 6 (211 -- 275 GHz), 7 (275 -- 373 GHz) and 9 (602 -- 720 GHz). See Table~\ref{table:bands} for a definition of all the ALMA bands\footnote{Adapted from http://www.almaobservatory.org}, which we will use throughout the paper for simplicity. The complete sample includes observations of 102 bursts. There have been 88 searches for GRB afterglows, of which 22 have reports of detections. Separating into individual bands, we have 18 detections in band 3, one in band 4, six in band 6, five in band 7 and none in band 9. Table~\ref{table:rat} displays a list of the detection ratios for each of the observing bands. As for the host galaxies, there have been specific host galaxy searches for 36 cases, although limits can be provided for the 102 bursts that have been followed. Host galaxy detections have only been achieved in four cases: GRB\,000210, GRB\,000418, GRB\,010222 and XT\,080109, this last one cannot be considered a normal GRB as the burst was in X-rays and not in gamma \citep{sod08}. There have been further claims of host detections but we consider only those with signal to noise ratio greater than 3.

\addtocounter{table}{1}

\begin{table}[h]
\caption{
Definition of the ALMA bands. The four marked in bold will be the main observing bands and the only ones offered for early science.
}            
\label{table:bands}      
\centering                          
\begin{tabular}{c c c}        
\hline\hline                 
Band 	& Frequency range 		& Wavelength range\\     
	 	& (GHz)		        		& (mm)	\\
\hline                        
Band 1	&  31 -- 45				&  6.66 -- 9.67	\\
Band 2	&  67 -- 90				& 3.33 -- 4.47	\\
\textbf{Band 3}	&  \textbf{84 -- 116}	& \textbf{2.58 -- 3.56}	\\
Band 4	& 125 --163			& 1.84 -- 2.40	\\
Band 5	& 162 -- 211			& 1.42 -- 1.85 	\\
\textbf{Band 6}	& \textbf{211 -- 275}	& \textbf{ 1.09 -- 1.42}	\\
\textbf{Band 7}	& \textbf{275 -- 373}	& \textbf{ 0.80 -- 1.09}	\\
Band 8	& 385 -- 500			& 0.60 -- 0.78	\\
\textbf{Band 9}	& \textbf{602 -- 720}	& \textbf{0.42 -- 0.50}	\\
Band 10	& 787 -- 950			& 0.32 -- 0.38	\\
\hline                                   
\end{tabular}
\end{table}

\begin{table}[h]
\caption{Detection ratios in the catalogue for each of the observing bands, considering only 3$\sigma$ or higher significance detections. For host galaxies we consider only dedicated searches and not the limits derived from afterglow searches.}             
\label{table:rat}      
\centering                          
\begin{tabular}{c c c}        
\hline\hline                 
 & Afterglows & Host galaxies \\    
\hline                        
Band 3	& 18 / 61 (30\%)	& 0 / 3 (0\%)	\\
Band 4	& 1 / 2 (50\%)		& ---		\\
Band 6	& 6 / 38 (16\%)		& 2 / 8 (25\%)	\\
Band 7	& 5 / 31 (16\%)		& 3 / 27 (11\%)	\\
Band 9	& 0 / 6 (0\%)		& 0 / 15 (0\%)	\\
Total		& 22 / 88 (25\%)  	& 4 / 36 (11\%)		\\
\hline                                  
\end{tabular}
\end{table}

Figure~\ref{lcs} shows a compilation of the mm/submm GRB afterglow light curves for ALMA bands 3, 6 and 7. Figure~\ref{hist} shows histograms of the peak detections and detection limits. The peak detections are, strictly speaking, only lower limits to the peak flux density in each band, as only a few light curves cover the maximum of the light curve. The detection limits are the single-epoch deepest 3$\sigma$ limits for each burst, independently of the time of the observation, so they can only be indicative of the capabilities of pre-ALMA observatories.

The afterglow peak flux for each observing band in the mm/submm range is expected to be very similar, with only the time at which the peak is reached being different. According to the fireball model \citep[][see aslo Sect.~\ref{sect:fb}]{sar98}, the peak is expected to cross from higher to lower energies, so that the higher frequency bands are expected to peak first. Hence, the ratio of afterglow detections is mostly determined by the different observing sensitivities of each of the telescopes in each band, given a reasonably prompt reaction time. Table~\ref{table:1st} displays the median time between the burst onset and the first observation for the bursts in the sample, together with the median 3$\sigma$ limiting flux of the earliest observations. We do not include Band 4, as it has only 2 observations and the statistics are not significant.

\begin{table}[h]
\caption{Median time between the burst and the first observation in each band and median limiting flux of those first observations (3$\sigma$).}             
\label{table:1st}      
\centering                          
\begin{tabular}{c c c}        
\hline\hline                 
 & Time & Limiting flux\\    
 & (day) 	& (mJy)		\\
\hline                        
Band 3	& 2.59 	& 0.87	\\
Band 6	& 2.67 	& 3.87	\\
Band 7	& 2.06 	& 7.50	\\
Band 9	& 4.47	& 60.0	\\
\hline                                   
\end{tabular}
\end{table}

   
\subsection{The physics of GRBs and their environments}
\label{sect:fb}

GRB afterglows can be described, in the simplest case, using the fireball model \citep{sar98}. 
According to this model, material is ejected at ultrarelativistic velocities through collimated 
jets (with opening angle $\theta_j$). When this material interacts with the medium surrounding the progenitor, the accelerated particles emit 
a synchrotron spectrum that is characterised by three break frequencies:
$\nu_m$ is the characteristic synchrotron frequency and is the maximum of the emission; 
$\nu_c$ is the cooling frequency, above which radiative cooling is significant;
$\nu_a$ is the synchrotron self-absorption frequency. 

\begin{figure}[h]
   \centering
   \includegraphics[width=9cm]{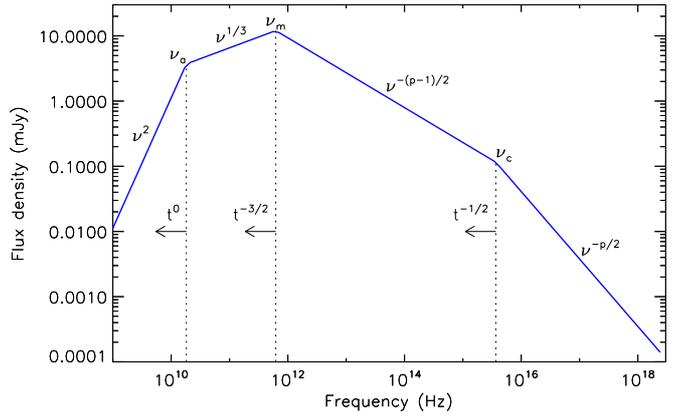}
      \caption{The synchrotron afterglow spectrum expected from a simple fireball model in the slow-cooling regime, assuming a constant density environment and spherical expansion. The figure shows the characteristic frequencies, their evolution and the spectral slopes of the spectrum \citep[adapted from][]{sar98}.
                    }
         \label{fig:fb}
   \end{figure}

The location of the spectral breaks at a given time, and the spectral slopes, shown in Fig.~\ref{fig:fb}, are determined by the following five parameters: $E$ -- isotropic equivalent kinetic energy of the shock; $n$ -- particle number density in the surroundings of the GRB, which can be considered constant or following a stellar wind profile; $\epsilon_B$ -- fraction of the shock energy that goes into magnetic energy density; $\epsilon_e$ -- fraction of the shock energy that goes to electron acceleration; $p$ -- slope of the relativistic electron energy distribution in the shock. Detailed multi-wavelength modelling of the spectral energy distribution allows us to derive these micro- and macro-physical parameters of the emission. 

   \begin{figure}[h]
   \centering
   \includegraphics[width=9cm]{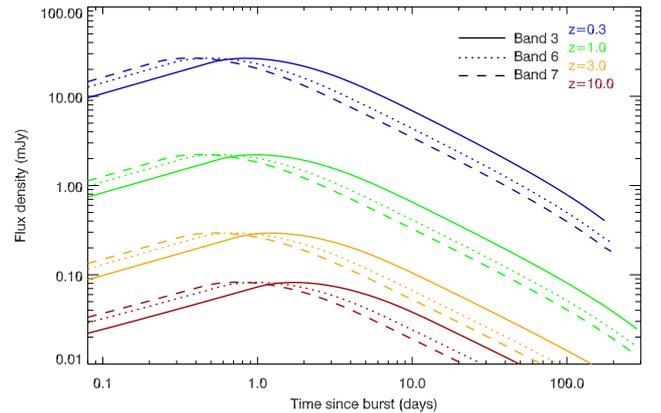}
      \caption{Light curves of what could be a typical GRB afterglow using the standard fireball model, plotted for the different ALMA bands and at different redshifts.
                    }
         \label{fig:modelc}
   \end{figure}

Figure~\ref{fig:modelc} shows an example of what we would expect the light curves of a typical GRB to look like in the different ALMA bands and at different redshifts using the standard fireball model. It assumes an homogeneous jet, adiabatic dynamics \citep[based on][]{rho99}, constant density environment, and integration over equal arrival time surfaces. The physical parameters of the fireball are the following: $E=10^{53}$ erg s$^{-1}$, n=0.1cm$^{-3}$, $\epsilon_B$=0.01, $\epsilon_e$=0.088, $p$=2.1 and $\theta_j$=5 deg. The light curves are characterised by a shallow rise during the first hours to days after the burst followed by a decay, as seen in most of the light curves of Fig.~\ref{lcs}, when there are enough data points. The figure also shows how the peak is reached later at lower frequencies for a given redshift, while it is delayed for high redshift events due to cosmological time dilation.

A reverse shock, produced inside the ejecta, can produce additional emission at early times \citep{pir99}. This emission has been rarely observed in the optical wavelengths \citep[e.g.][]{ake99,bla05,jel06,rac08,zhe11} but is expected to have a significant contribution in the early mm/submm emission \citep{ino07}. For example, the mm detection of GRB\,090423, at a redshift of 8.2, seems to show excess emission possibly due to a reverse shock \citep{cas09b,cha10}.

This model is, however, known to be too simplistic to explain the complex evolution that we can see in the densely sampled X-ray and optical light curves of \emph{Swift} GRBs. Several modifications have been suggested to explain some of the observed fluctuations, flares, bumps and wiggles. Such modifications may include different density profiles and fluctuations \citep{wan00,ram01,dai02,nak03}, energy injections \citep{ree98,sar00,gra03,bjo04,joh06}, jets with complex structure \citep{mes98,kum00,ros02}, late engine activity \citep{dai98,zha02,ram04}, microlensing \citep{loe98,gar00,iok01} or dust echoes \citep{esi00,mes00,rei01}. Discerning the different scenarios is only possible through the analysis of the SEDs obtained with multi-wavelength observations (See Sect.~4) and, in some cases, with the aid of polarimetry.

\subsection{Redshift distribution and luminosities}

\begin{figure}[h!]
   \centering
   \includegraphics[width=9cm]{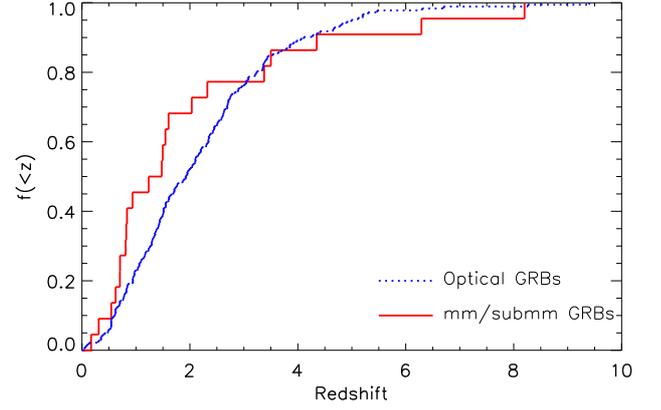}
      \caption{The cumulative fraction of GRBs as a function of redshift for the optical sample of \emph{Swift} GRBs \citep[dotted curve; ][]{jak11} and mm/submm-detected bursts (solid curve).}
         \label{redshift}
   \end{figure}

In spite of the strong limitation imposed by the detection limits of current observatories, the sample includes detections of GRBs at most redshifts, ranging from $z$=0.168 (GRB\,030329 \citealt{gre03}) all the way up to $z$=8.2 (GRB\,090423 \citealt{tan09,sal09}). XT\,080109 at a redshift of 0.0065 \citep{mal08} had no gamma-ray emission, so we will exclude it from the analysis. This shows that GRBs can be great tools to study the evolution of the star formation across the complete history of the Universe also in this frequency range. Figure~\ref{redshift} shows the distribution of the redshifts of GRBs detected in the mm/submm range as compared to the distribution obtained from optical observations \citep{jak11}. The mm/submm sample has a larger percentage of low redshift detections but, on the other hand, the average redshift was compensated by deep searches of the high redshift events.
We note that the optical sample has its own biases, for example the limited capability to detect dark bursts (see Sect. 4.1 for a definition of dark burst). A Kolmogorov-Smirnov test gives a probability of 11\% for both mm/submm detected bursts and the \emph{Swift} sample of GRBs with redshift of coming from the same population. We obtain a median (mean) redshift for our sample of detections of 1.48 (1.99), as compared to 1.92 (2.19) of the sample of \emph{Swift} bursts \citep{jak11}.

   \begin{figure}[h!]
   \centering
   \includegraphics[width=9cm]{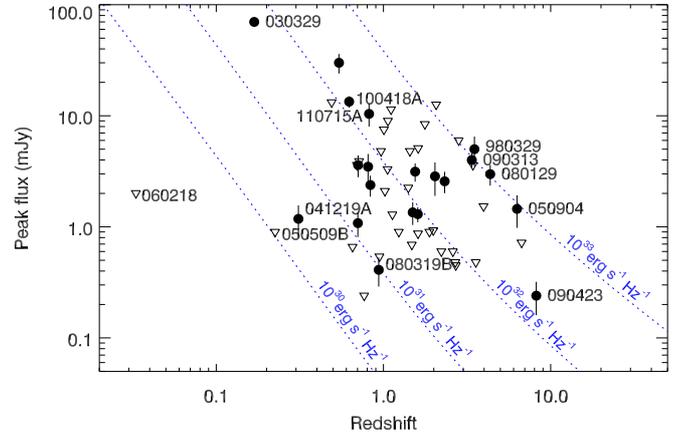}
      \caption{Peak flux density measured in mm/submm (as in Fig.\ref{hist}) vs. redshift. Filled dots indicate detections whereas empty triangles are detection limits. The dotted lines indicate the flux density levels for equal luminosity objects at varying redshifts. Some interesting bursts are indicated in the figure.}
         \label{fig:peak_z}
   \end{figure}
   
   In Fig.~\ref{fig:peak_z} we show the peak flux density measured in the mm/submm range as a function of the redshift. The dotted lines indicate the expected flux at different redshifts for objects of equal peak luminosities. The figure shows that the brightest burst ever detected, GRB\,030329 \citep{hjo03,sta03,lip04,she03,res05}, is mainly so due to its low redshift. GRB\,090423, the furthermost detected in mm/submm, at a redshift of $z=8.2$ \citep{sal09,tan09}, was not extremely luminous, and was only detected thanks to very deep observations (it is, in fact, the deepest detection in the sample; \citealt{cas11}). GRB\,080319B, also known as the ``naked-eye burst'' due to its extreme optical brightness \citep{rac08,blo09,pan09} is found in the faint end of the distribution, indicating that most of the emission was concentrated in the optical bands.
      
   On the upper part of the diagram, we identify a family of very luminous events, peaking at $\sim10^{33} \rm{erg}$ $\rm{s}^{-1} \rm{Hz}^{-1}$, composed by GRB\,980329 \citep{smi99,jau03}, GRB\,050904 \citep{tag05,hai06}, GRB\,080129 \citep{gre09} and GRB\,090313 \citep{mel10}, all beyond redshift 3. If those bursts were to be found at redshift of $\sim1$ they would reach peak fluxes of 40 mJy and easily over 100 mJy at redshifts of $\sim0.5$. 
   
   On the fainter side, GRB\,060218 ($z=0.03$) is be the burst for which we would have the most constraining peak luminosity limits. However, this burst is known to be peculiar, being its optical emission dominated by a supernova component and not the afterglow (\citealt{cam06,sod06}; see also \citealt{tho11}). GRB\,050509B an extremely faint short burst, most probably hosted by a giant elliptical galaxy at $z=0.22$ \citep{hjo05,geh05,cas05} is the following dimmest limit. The least luminous burst with detection in the mm/submm range is GRB\,041219A, which curiously was one of the longest and brightest GRBs detected, for which a redshift of $z=0.31$ has been recently suggested \citep{got11}.
   
   In the figure we also identify GRB\,100418A and GRB\,110715A, whose counterparts were discovered with our programmes. They are the 3rd and 4th brightest bursts ever detected in the mm/submm bands.
   
   Using all the detections, we obtain an average peak spectral luminosity of $10^{(32.1\pm0.7)} \rm{erg}$ $\rm{s}^{-1} \rm{Hz}^{-1}$. This is roughly an order of magnitude brighter than the value found by \citet{cha11} in cm wavelengths, as expected by afterglow models. However, we note that these values must be used with care, as both samples strongly biased due to a low detection rate (25\% in our case and 30\% in the case of the cm range study).

   \begin{figure}[h]
   \centering
   \includegraphics[width=9cm]{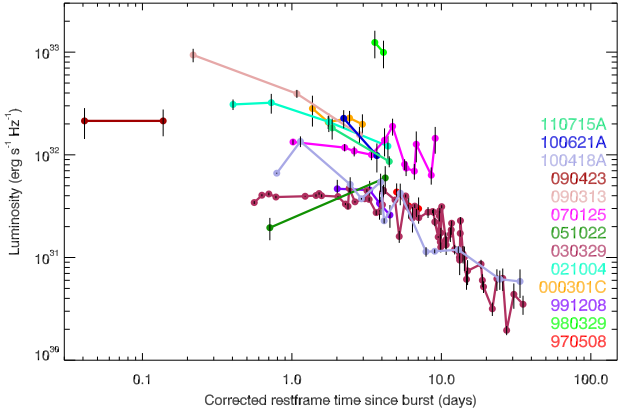}
      \caption{Afterglow light curves transformed to the rest frame of the central engine. The data points of the different bands have been shifted using an offset based on the fireball model, as explained in the text.
                    }
         \label{fig:lcrest}
   \end{figure}

Finally, we collect the light curves of GRBs with at least 2 detections and a well determined redshift, and transform them to the rest frame of the central engine to compare them directly. In Fig.~\ref{fig:lcrest} we plot the light curves (only detections) in luminosity and corrected of cosmological time dilation. To be able to use all the bands together, we introduce a temporal offset assuming that the light curves are described by a fireball model in which the peak frequency is evolving with $t^{-3/2}$ (see Sect.~\ref{sect:fb}). This is a simplistic approach but enough for a qualitative view. The intrinsic light curves appear much more clustered than the observed ones in Fig.~\ref{lcs} where the dispersion in fluxes was mostly due to redshift: The brightest light curve was from GRB\,030329, the nearest burst and the faintest from GRB\,090423, the furthest one, whereas in Fig.~\ref{fig:lcrest} they are both normal luminosity bursts. On the other hand, the light curves appear to be reasonably well described by the expectations of the theory described in Sect.~\ref{sect:fb}, reaching a maximum between a few hours and a few days and then decaying.

\subsection{Estimation of the peak flux density distribution}
\label{sec:flxdens}

\begin{figure}[h!]
   \centering
   \includegraphics[width=9cm]{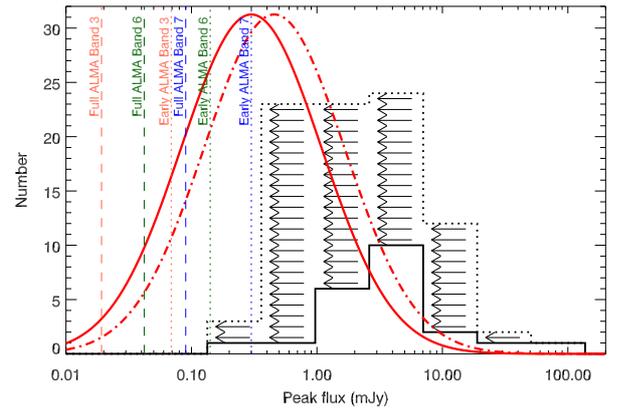}
      \caption{Histogram showing the detections and detection limits of all the bands combined and an estimate of the real peak flux density distribution (indicated with a thick trace), compared to the detection limits of ALMA (from Table~\ref{table:rms}). The dashed-dotted gaussian is the distribution derived from our sample, before correcting for the selection bias, as explained in the text.
                    }
         \label{fig:estimate}
   \end{figure}

We can try to make a rough estimate of the distribution of peak flux densities using the data from our sample and information extrapolated from other spectral ranges. We assume that the peak flux density of GRB afterglows can be described by a gaussian function in the logarithmic space as is seen for the flux density at specific times in other wavelengths. This gaussian can be normalised using the fact that in band 3 we are detecting 30\% of the afterglows down to a limit of $\sim0.9$mJy. The largest uncertainty here comes from the dispersion of the gaussian. Using the complete catalogue of data from \citet{kan10} we find that the dispersion in the brightness of observed optical {\it R}-band magnitudes, corrected of extinction, at a specific time (1.0 days) is of 0.49, whereas for the X-rays, using the data from the burst analyser \citep{eva10} find a dispersion of 0.56. Both values are quite similar, however, as the X-ray constitutes a more complete sample we will use this estimate. We must also consider that the sample of mm/submm follow-up mainly includes brighter bursts. If we calculate the average X-ray flux densities at one day of all the GRBs in the \emph{Swift} sample and compare it with the subsample of the ones that were followed in mm/submm we find that the one first is dimmer by a factor of 1.7, which we apply to the average of the estimated distribution in mm/submm.
Figure~\ref{fig:estimate} shows a histogram with the detections and detection limits of all bands combined, together with the distribution of peak fluxes that has been estimated here, with an average of 0.3 mJy. Using these values we estimate that ALMA should be able to detect, using band 3, 87 \% of all the afterglows already in its early configuration with 16 antennas, and 98 \% in its full setup of 66. Using band 6, the values would be 72 \% (early) and 94 \% (full), and in band 7, 50 \% (early) and 82 \% (full).

\subsection{Comparison with X-ray flux densities}

Thanks to the X-Ray Telescope \citep[XRT; ][]{bur05} onboard \emph{Swift}, nowadays we have early X-ray detections of the vast majority of GRB afterglows within the first minutes after the burst. These observations give us not only a precise position of the afterglow, but also a spectrum from which we can derive the unextinguished flux density, the spectral slope and the extinction due to metals in the line of sight. All this information can be used to estimate the expected flux in other wavelengths on a case to case basis assuming a synchrotron model as described in Sect.~\ref{sect:fb}.

\begin{figure}[h]
   \centering
   \includegraphics[width=9cm]{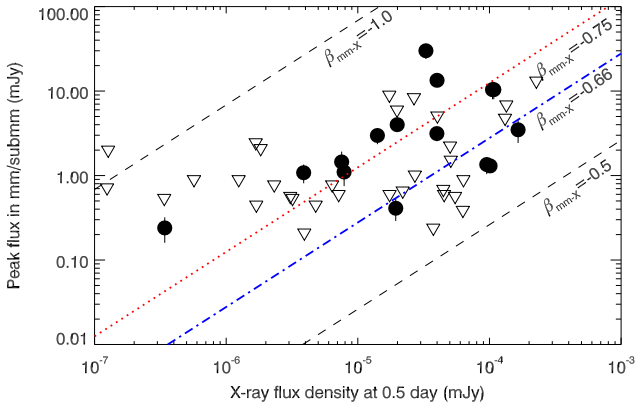}
      \caption{Comparison of the peak flux density measured in mm/submm with the X-ray flux density 0.5 days after the burst (at 10keV). Filled dots indicate detections, whereas empty triangles are 3$\sigma$ detection limits. Dashed lines indicate some characteristic slopes of the fireball model. The dotted line is the average observed in the sample of detections and the dashed-dotted line is an estimation of the real average.
                    }
         \label{fig:Xmm}
   \end{figure}

In this section we make a general comparison of the flux densities measured in the X-ray and mm/submm bands for those bursts that were observed by \emph{Swift}. For the X-rays, we use the flux densities measured 0.5 days after the burst (when we can assume that the emission is dominated by the afterglow) at 10 keV \citep{eva10}. The scarce and heterogenous mm/submm coverage does not allow us to obtain a sample of flux densities at a given epoch. In consequence, we use, as an approximation the peak flux densities, as explained in Sect.~\ref{sec:pre}. The two quantities are plotted in Fig.~\ref{fig:Xmm}, where we can see, as expected, a correlation between the flux density measured in X-rays and in mm/submm, although with a significant dispersion.

\citet{eva09} found that the typical spectral slope measured in X-rays was, on average $\beta_X\sim-1.0$ (where $F_{\nu}\propto\nu^\beta$), which would be the slope expected for an electron index of $p=2.0$ if X-rays were above the cooling break ($\nu_c$). The extrapolation of the X-ray measurement to mm/submm using this slope would give us the maximum flux that we could expect to measure for a synchrotron spectrum and is indicated in the figure by the upper dashed line. As expected, there is no GRB that has been measured above this extrapolation. On the other side, if $\nu_c$ was just below X-rays, the lowest extrapolated flux would be obtained using a spectral index of $\beta=-0.5$, also indicated in the figure.  This is what \citet{jak04} used to define the limit for dark bursts (see Sect.~\ref{sec:ala}). In the case of mm/submm, due to the lack of extinction or due to dust or high redshift, we would only expect to find below this threshold those bursts where $\nu_m$ would be above the mm/submm range. In Fig.~\ref{fig:Xmm} we can see that all our detected bursts are located between these two limits, and mostly clustered around $\beta=-0.75$. On the other hand, the average flux density derived from X-rays 0.5 days after the burst (derived in a similar way as we did for 1 day in Sect.~\ref{sec:flxdens}) at 10keV is $1.04\times10^{-5}$ mJy, whereas we estimated at typical peak flux density of 0.3 mJy. This would imply a typical $\beta_{mm-X}$=0.65, slightly lower than the typical value for detected bursts, as expected due to the abundant detection limits.


\section{GRBs in the ALMA era}

\begin{table*}[]
\caption{Continuum flux density sensitivity (1$\sigma$) with 1 hr on-source observations. Based on the exposure time calculators of the different observatories operating in the mm/submm range. We assumed a precipitable water vapour of 1mm and an average elevation of the source of 60 deg. }             
\label{table:rms}      
\centering                          
\begin{tabular}{c c c c c}        
\hline\hline                
Obs./Inst. 							& Band 3 	& Band 6	& Band 7 	& Band 9	\\     
								& (mJy)	& (mJy)	& (mJy)	& (mJy)		\\
\hline                        
APEX (LABOCA/SABOCA)\tablefootmark{1,2}		&	---	&	---	&	2.6	&	22.2	\\
SMA\tablefootmark{3}				&	---	&	0.8	&	2.0	&	29	\\
IRAM30m (MAMBO2)\tablefootmark{4}	&	---	&	0.76	&	---	&	---	\\
PdBI (WIDEX)\tablefootmark{5}			&	0.08	&	0.18	&	---	&	---	\\
JCMT (SCUBA2)\tablefootmark{6}		&	---	&	---	&	0.9	&	3.6	\\
CARMA\tablefootmark{7}				&	0.43	&	0.68	&	---	&	---	\\
\hline
ALMA (Early)\tablefootmark{8}			& 2.3$\times10^{-2}$	& 4.7$\times10^{-2}$	& 0.10				& 0.72	\\
ALMA (Full)\tablefootmark{8}			& 6.4$\times10^{-3}$	& 1.4$\times10^{-2}$	& 3.0$\times10^{-2}$	& 0.20	\\
\hline                                   
\end{tabular}
\begin{center}
\tablefoottext{1}{http://www.apex-telescope.org/bolometer/laboca/obscalc/}\\
\tablefoottext{2}{http://www.apex-telescope.org/bolometer/saboca/obscalc/}\\
\tablefoottext{3}{http://sma1.sma.hawaii.edu/beamcalc.html}\\
\tablefoottext{4}{https://mrt-lx3.iram.es/nte/time\_estimator.psp\#mambo}\\
\tablefoottext{5}{http://www.iram.fr/IRAMFR/GILDAS/}\\
\tablefoottext{6}{http://www.jach.hawaii.edu/JCMT/continuum/scuba2\_integration\_time\_calc.html}\\
\tablefoottext{7}{http://bima.astro.umd.edu/carma/observing/tools/rms.html}\\
\tablefoottext{8}{http://www.eso.org/sci/facilities/alma/observing/tools/etc/}\\
\end{center}
\end{table*}

As shown in Figs.~\ref{lcs} and \ref{hist}, until now the mm/submm observations of GRBs have been limited by the low sensitivity of the observations in this range as compared to other wavelengths. This is changing with the start of operations of the ALMA observatory. Even in its early configuration, ALMA will outperform the sensitivity of all previous observatories. Table~\ref{table:rms} shows a comparison of the sensitivities (assuming equal observing conditions) of some of the current mm/submm observatories with ALMA, both in its early and final configuration. In its final setup, the sensitivities will improve between 1 and 2 orders of magnitude. If we consider that the atmospheric conditions at the ALMA site are commonly better than the average of other observatories, the difference will be even more significant.

However, it will not only be the sensitivity, but also the spatial resolution that will allow ALMA to make breakthroughs in GRB studies. In its extended configuration, ALMA in its full array capabilities will allow resolutions of 0.050$^{\prime\prime}$ in band 3 or 0.013$^{\prime\prime}$ in band 7 (4.8$^{\prime\prime}$ and 1.7$^{\prime\prime}$, respectively, in the most compact configuration), allowing precise localisation of the afterglows and detailed studies of the host galaxies.

The observation of GRB\,110715A that we present here, obtained during the commissioning phase of ALMA, is an advance of what this observatory will be capable of. Already the commissioning phase it was proved that ALMA can reliably perform ToO observations. Furthermore, with only 25 minutes on target, it obtained the deepest observation of a GRB in the 345 GHz band to date and allowed us to derive the most precise coordinates that are available for this afterglow.

\subsection{GRB afterglows}
\label{sec:ala}

ALMA observations will play a significant role in understanding the physics of GRBs and their environment. The increased sensitivity will allow us to probe almost complete samples of afterglows, not just limited to the brightest events as we do now, eliminating most of the observational biases. In the case of very bright events ALMA will allow us to attempt, for the first time, studies of absorption spectral features, from which we will be able to derive the molecular content of distant galaxies and the chemical enrichment of the interstellar medium along the history of the Universe.

   \begin{figure}[h]
   \centering
   \includegraphics[width=9cm]{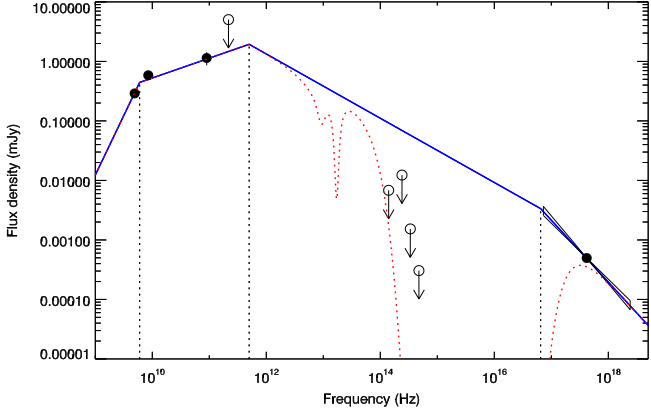}
      \caption{SED of the afterglow of the dark GRB\,051022. Filled dots mark detections while empty dots and arrow mark detection limits. In blue we draw a synchrotron spectrum that can explain the radio and X-ray emission, while the optical/nIR data ($10^{14-15}$ Hz) lie well bellow. Adding a Small Magellanic Cloud (SMC) extinction of $A_V \sim 15$ results in a SED (red dotted line) consistent with all the data \citep[data obtained from][]{cam05,bre05,cas07}.
                    }
         \label{fig:dark}
   \end{figure}

   \begin{figure}
   \centering
   \includegraphics[width=9cm]{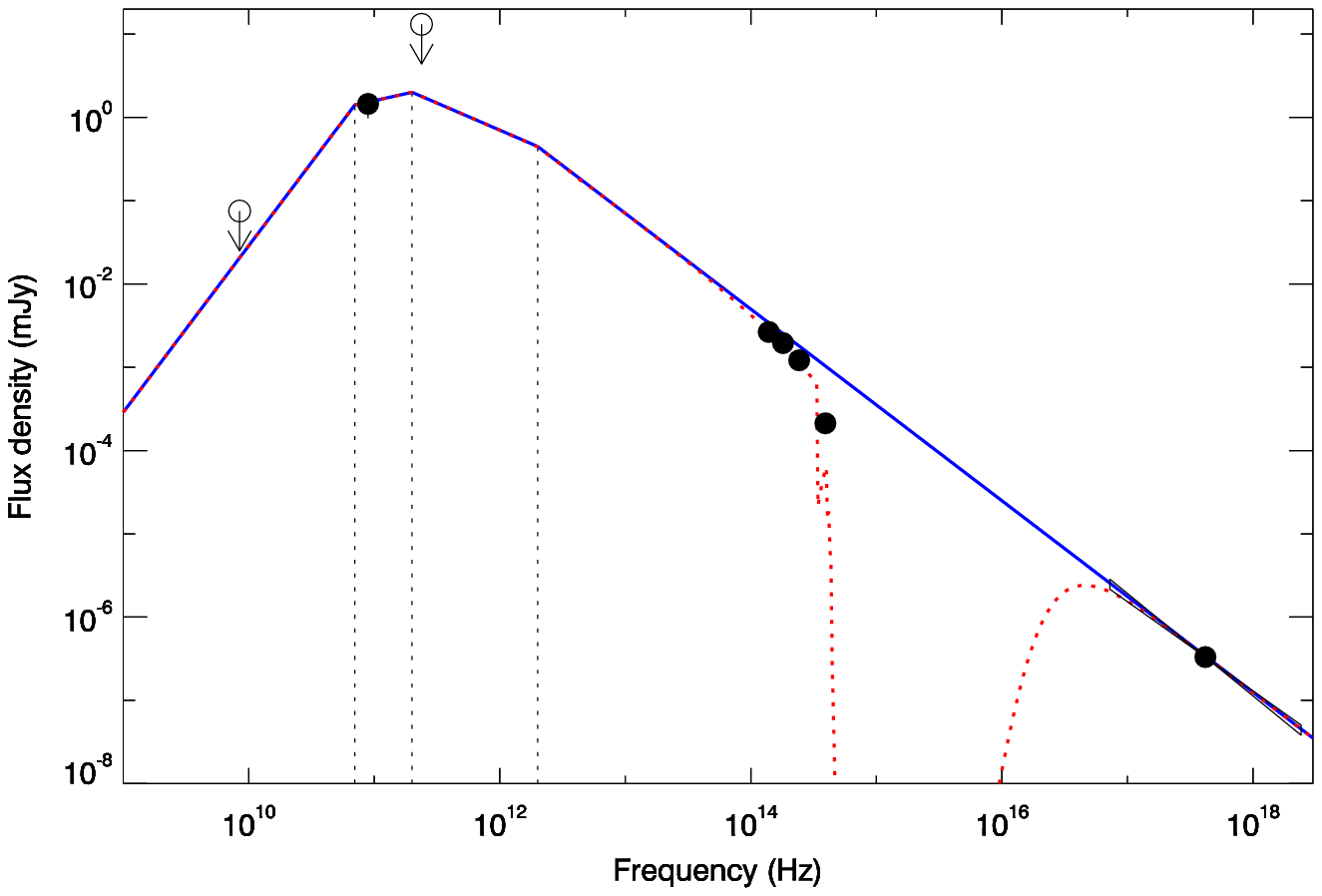}
      \caption{SED of GRB\,050904 at redshift $z=6.29$ \citep{kaw06} using the detection from PdBI 5.896 days after the burst \citep{cas11}. The X-ray flux is estimated from the XRT light curve, the radio data are from \citet{fra06}, while optical/NIR are extrapolations from \citet{tag05,hai06}. For the rough synchrotron spectrum drawn in blue, we have assumed an electron energy distribution index of $p=2.30$, while an SMC extinction with $A_V=0.12$ (consistent with the value derived by \citealt{str11}; but see also \citealt{zaf10})  and Lyman blanketing \citep{mad95} have been considered in the red dotted line to match the optical/NIR observations. With these data we can constrain the position of $\nu_a<2\times10^{11}$ Hz and $2\times10^{11}<\nu_c<10^{13}$ Hz.
                    }
         \label{fig:hiz}
   \end{figure}

Thanks to its large collecting area, ALMA will also be able to provide polarimetric measurements for a sample or relatively bright events. As shown by \citet{tom08}, this can give a unique view on some fundamental physics of shockwaves (though ALMA polarimetric calibration is unlikely to get better than $\sim$ 1 percent). Furthermore, simple linear polarisation measurements in submm and optical, will tell us about the plasma properties in the forward shock and the geometry of the ejecta.

One of the key areas where ALMA will play an important role will be the study of dark bursts, which strongly bias the optically selected samples of GRBs \citep[see for example][]{fyn09}. Dark bursts are those that are less luminous in the optical than what we would expect them to be assuming a synchrotron spectrum as the one explained in Sect.~\ref{sect:fb} based on the extrapolation of X-ray observations. \citet{jak04} define them assuming that the electron index is $p\geqslant2$ \citep{sar98}, and that $\nu_m$ is below optical frequencies (as is normally the case). Under this assumption, we would never expect a synchrotron spectrum with a slope between optical and X-rays $\beta_{{\rm OX}} > -0.5$ (note that in this paper we use the convention $F_{\nu}\propto\nu^\beta$ as in \citealt{sar98}, $\beta$ having opposite sign to the one used by \citealt{jak04}). This definition was extended by \citet{van09} not limiting the value of $p$ and, instead, requiring $\beta_{{\rm OX}} > \beta_{\rm X} + 0.5$ (being $\beta_{\rm X}$ the spectral slope measured in X-rays), given that the X-ray spectral slope is normally well constrained. This reduced optical emission can be due to optical extinction by dust, a high-redshift event (where the Lyman break is beyond the optical, at $z \gtrsim 5$) or a deviation from the simple synchrotron spectrum (like an additional inverse Compton component or a distribution of the emitting electrons not described by a single power law). Below we give an example of each of the two main families of dark bursts.

   \begin{figure*}
   \centering
   \includegraphics[width=15cm]{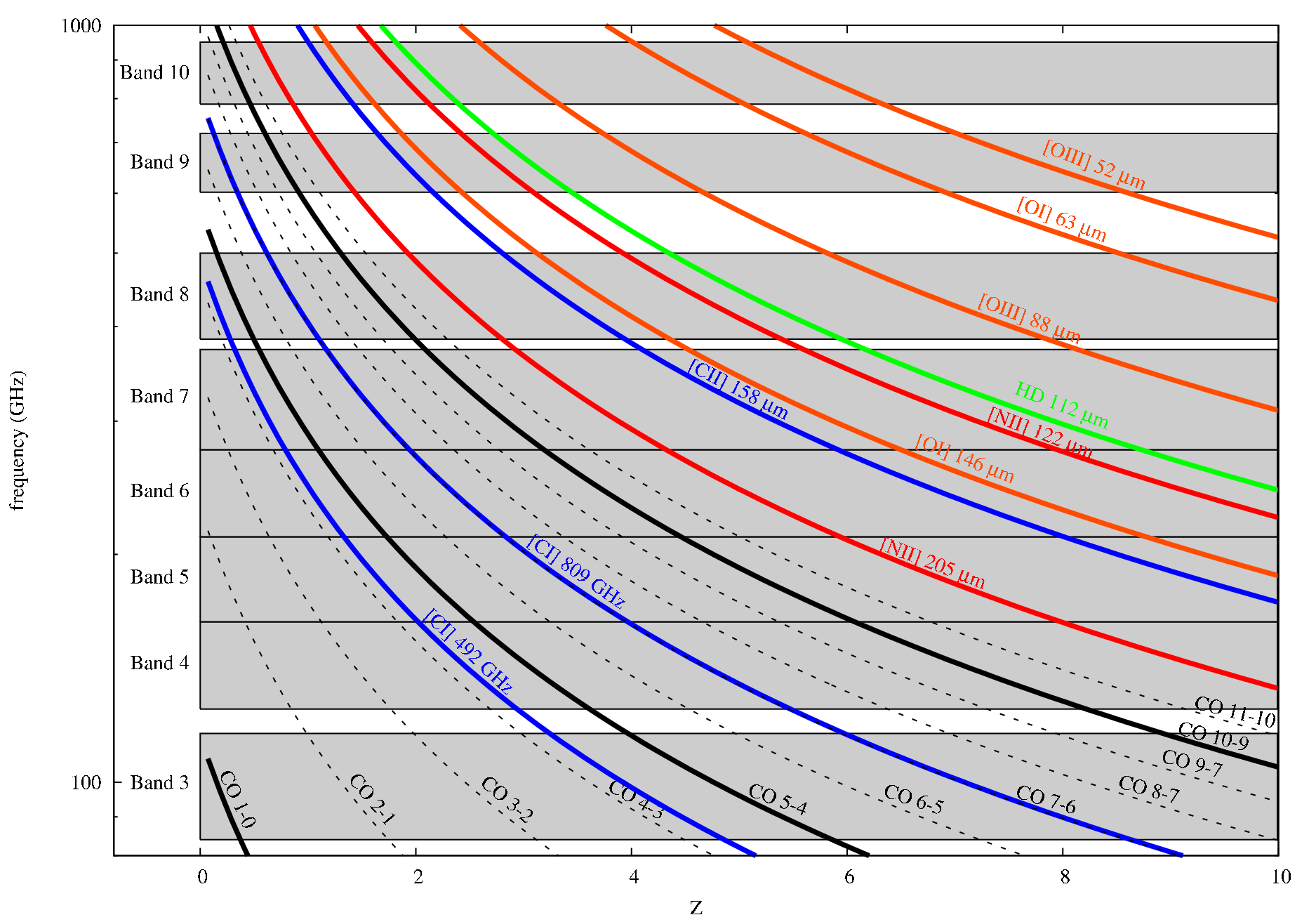}
      \caption{Spectral lines observable at each ALMA band as a function of the redshift. Different colours have been used to indicate different elements.
                    }
         \label{fig:lines}
   \end{figure*}

Optically selected samples of GRBs are limited by the amount of extinction in the host galaxy that, if large, can attenuate the optical emission and make it undetectable. Hence, there is a strong bias against high extinction bursts, which limits our capability to study the star forming regions where GRBs are produced. However, the negligible effect of dust extinction in the mm/submm range allows us to detect afterglows independently of this effect, and consequently study a more complete sample. As an example of highly extinguished burst, we can look at the case of GRB\,051022, with $\beta_{{\rm OX}} > 0.05$ \citep{cas07,rol07} one of the darkest bursts detected to date, for which an optical counterpart was not found. Precise location was obtained thanks to mm observations \citep{cam05,bre05,cas07}, which allowed identification of the host galaxy, for which a redshift of $z=0.8$ was derived \citep{cas07}. A broadband study allows us to impose a lower limit on the host galaxy extinction of $A_V>15$ (see Fig.~\ref{fig:dark}, and for other estimates also \citealt{cas07,rol07}), far beyond what is typically seen in optically selected GRB afterglows.

The other main cause for optically dark GRBs is a high redshift. In these cases the absorption produced at frequencies higher than the Lyman limit does not allow us to obtain optical detections (in $R$-band) of GRB afterglows beyond redshifts of 6. These events are important to understand the formation of the first stars in the Universe. Proof that they can be detected in the mm/submm range is the fact that already in the pre-ALMA era, out of the three GRB afterglows observed at $z>6$, two have been detected (GRB\,050904 at $z=6.3$ and GRB\,090423 at $z=8.2$). In Fig.~\ref{fig:hiz} we have drawn the SED of GRB\,050904 at $z=6.3$, including radio limits \citep{fra06}, a detection in band 3 (and a limit in band 6) from Plateau de Bure Interferometer \citep{cas11}, near-infrared and optical data \citep{tag05,hai06} and X-ray data from \emph{Swift}. In spite of its high redshift, this burst was detected with a peak flux density of 1.47 mJy in band 3, 80 times above ALMA's detection threshold for 1 hr of observations in band 3. In the case of bright events it has been suggested that the study of the HD deuterium-molecule rotational-transition in absorption at 112 $\mu$m could give us important clues on the formation of population III stars at very high redshift \citep{ino07}. CO absorption lines, which have been already reported through optical spectroscopy of the afterglow of GRB\,080607 \citep{pro09,she09} are also good candidates to produce absorption features in the mm/submm range.

\subsection{GRB host galaxies with ALMA}

Up to now the majority of GRB hosts has not been detected at mm/submm  wavelengths  \citep{tan04,ber03,priddey06,wat10} even when dark GRBs (which are supposed to be dusty) were targeted \citep{bar03}. This was interpreted as an indication that they are not heavily dust-obscured. Except for the host of the local GRB 980425 (Micha{\l}owski et al.~in preparation), only three $z\sim1$ GRB hosts (GRB\,000210, GRB\,000418 and GRB\,010222) were detected at mm and submm wavelengths  \citep{fra02,ber03,tan04}. Their derived dust masses of a few times $10^8\,M_\odot$ \citep{michalowski08} are comparable to that of $z\sim2$ -- $3$ dusty starbursts known as submm galaxies \citep{kovacs06,michalowski10smg}.  Despite these significant amounts of dust,  these GRB hosts exhibit properties consistent with the rest of the GRB host population, namely blue optical colours, low extinction and low optical star formation rates \citep{gorosabel1, gorosabel2,galama, ber03,savaglio03,savaglio09,christensen04,castroceron06,castroceron10,kann06}. This discrepancy was explained by \citet{michalowski08} by invoking very young stellar population with optical emission completely extinguished by dust. This indicates that it is virtually impossible to predict the dust emission of GRB hosts based on their optical properties and advocates for the use of ALMA to target a significant and unbiased sample of these galaxies.

Despite numerous attempts, CO emission has not been detected for any GRB host galaxy \citep{koh05,endo07,hatsukade07,hatsukade11b, stanway11}; we refer to Table 1 in \citet{hatsukade11b} for a homogeneous overview of all available limits on CO line luminosities of GRB hosts. The derived upper limits on molecular gas masses of $10^8$ -- $10^9\,M_\odot$ indicate that GRB hosts are not as gas-rich as submm galaxies \citep{greve05,ivison11}.

We further note that the recent discovery of strong CO and H$_2$ features in absorption in an optical spectrum of the afterglow of GRB\,080607 \citep{pro09,she09} highlights the value of ALMA molecular line observations in tandem with optical spectroscopy. The detection of vibrationally excited transitions (H$^*_2$) in the afterglow of GRB\,080607 \citep{she09}, likely excited by the UV flux of the afterglow, show the importance of repeated spectroscopic observations by ALMA to probe the ISM properties of the host galaxies through time variable lines.

The unprecedented sensitivity of ALMA in the mm and sub-mm wavelengths, will allow the study of GRB host galaxies through both continuum and emission line observations. ALMA is sensitive enough to probe the dust properties of representative GRB hosts, not only the most star-forming ones. Moreover, the excellent angular resolution of ALMA will allow spatially-resolved studies of these galaxies.
This is important as there are indications that the immediate environments ($1$ -- $3$ kpc) of GRBs are the most star-forming sites within their host galaxies \citep{bloom02,fruchter06,thone08,ostlin08,michalowski09}. 

To date, measured spectroscopic GRB redshifts range between 0.0085 and 8.2.
Figure~\ref{fig:lines} shows a number of spectral features that will be covered by the different ALMA bands for a range of redshifts between $z=0$ up to $z=10$.
Taking into account the weather constraints at the highest bands, we consider that bands 3 to 7 will be better suited for host detections.
For the nearer events ($z\lesssim4$) the carbon monoxide ladder of rotational transitions up to $J=11$ -- $10$ could be feasible as reported 
for the submillimeter galaxy SMM J16359+6612 \citep[at $z=2.52$;][]{Weiss2005} and the broad absorption line quasar APM08279+5255 \citep[at $z=3.87$;][]{Weiss2007}.
Additionally, the fine structure transitions of neutral carbon at 492 and 809~GHz were also detected towards APM08279+5255 \citep{Weiss2005a,Wagg2006}.
Moreover, [CI] at 809~GHz and CO $J=7$ -- $6$, separated $\sim2.4$~GHz in the rest frame, could be observed simultaneously for redshifts of $z\sim2$ -- $8$ provided the large
bandwidth of ALMA receivers.
At redshifts above $z=4$ -- $5$, the higher-$J$ CO transitions will still be visible mostly within bands 3 to 5.
Above that redshift, the fine structure lines of the most abundant elements, O, C, and N, will be the best suited ISM tracers.
In particular the fine structure line of [CII] at $158\,\mu$m is likely the brightest emission line observable at high-$z$ \citep{Maiolino2009}.
The lines of [OIII] at 52~$\mu$m and 88~$\mu$m and [OI] at 63~$\mu$m have also been reported towards high-redshift lensed systems \citep{Ferkinhoff2010,Sturm2010}.


\section{Conclusions}

This paper presents the most complete catalogue of GRB observations in the mm/submm range to date, including observations of 11 GRBs from our programmes at APEX and SMA. We include a short observation of a GRB obtained with the ALMA observatory during its commissioning phase and with only 7 antennas, that gives an early idea of the potential of the observatory. The total sample contains data from 102 GRBs. Among this collection of observations, there are 88 observations of GRB afterglows, of which 22 (25\%) were detected (two of them discovered within our programmes) and 35 of host galaxies, of which 4 were detected. Observations until now have been strongly limited by the sensitivity of the instruments in this range, imposing important biases to any statistical study performed with them. Even with these limitations, we have proved that it is already possible to detect counterparts to GRBs in the mm/submm range with redshifts ranging from 0.168 all the way up to 8.2. This spectral range has also proved to be a very useful tool to localise and characterise the counterparts of dark GRBs, which are elusive in optical wavelengths.

Within the sample bursts have peak luminosities spanning over 2.5 orders of magnitude, with the most luminous reaching $10^{33} \rm{erg}$ $\rm{s}^{-1} \rm{Hz}^{-1}$. If such a burst happened at redshift of $\sim1$ they would reach peak fluxes of 40 mJy and over 100 mJy at redshifts of $\sim0.5$. Comparing with X-rays, we find a  correlation of the peak flux density in mm/submm with the X-ray flux density at 0.5 days that can serve to make rough predictions of mm/submm brightness from the early X-ray data. Using data from the sample and assumptions based on samples at other wavelengths, an estimate of the real peak flux density distribution of GRBs is made, from which an average peak flux density value of 0.33 mJy can be expected.

On the 30$^{\rm{th}}$ September 2011, the ALMA observatory started scientific operations with a limited setup of 16 antennas. Already then, it became the most sensitive observatory in the mm/submm range, increasing the sensitivity and resolution by around an order of magnitude with respect to the previous instrumentation, being even more significant in the highest frequencies. The final full observatory, with 66 antennas will set completely new standards in the field. In addition to the improved sensitivity and resolution, ALMA will be capable of observing in a very wide range of frequencies, that was never covered before by a single observatory. Using our peak flux density estimates and the detection limits calculated for ALMA, we can expect that early ALMA will detect around 87 \% of the bursts, while full ALMA should be able to detect 98 \%. In the case of bright GRB afterglows, like GRB\,030329, GRB\,100621A, GRB\,100418A or GRB\,110715A, where fluxes reached tens of mJy, ALMA will be able to study spectral features and perform polarimetric studies of the afterglow, which have been out of reach until now.

With ALMA we will be, for the first time, in position to undertake studies of samples of GRB host galaxies. We will be able to perform studies of the continuum emission to characterise the dust content and determine the unextinguished star formation rate of the hosts. Through the study of emission features from the host, we will be able to understand the molecular content and the chemical enrichment of the strong star-forming regions in which GRBs are found, at redshifts that go back to the epoch in which the first stars were formed.


\begin{acknowledgements}
AdUP acknowledges support from ESO's Scientific Visitor Programme in Chile, where a significant part of this work was done.
The Dark Cosmology Centre is funded by the Danish National Research Foundation.
The Atacama Large Millimeter/submillimeter Array (ALMA), an international astronomy facility, is a partnership of Europe, North America and East Asia in cooperation with the Republic of Chile. This paper makes use of the following ALMA commissioning data set: 2011.0.99001.CSV. We are grateful for the support from the ALMA commissioning team.
APEX is a collaboration between the Max-Planck-Institut fur Radioastronomie, the European Southern Observatory, and the Onsala Space Observatory. 
The Submillimeter Array is a joint project between the Smithsonian Astrophysical Observatory and the Academia Sinica Institute of Astronomy and Astrophysics and is funded by the Smithsonian Institution and the Academia Sinica. This paper made use of observations carried out with the IRAM Plateau de Bure Interferometer. IRAM is supported by INSU/CNRS (France), MPG
(Germany) and IGN (Spain). This work made use of data supplied by the UK Swift Science Data Centre at the University of Leicester.
We acknowledge partial support from the Spanish Ministry of Science and Innovation through programmes AYA20082008-03467/ESP and AYA2009-14000-C03-01. SS acknowledges support by a Grant of Excellence from the Icelandic Research Fund. 
\end{acknowledgements}


\bibliographystyle{aa}
\bibliography{almaGRB}


\Online

\begin{appendix}

\onllongtab{1}{
\begin{longtable}{l c c c c c c}       
\caption{Pre-ALMA observations of GRBs in the millimeter and submillimeter range. Epochs are given as mid-observing time since the GRB onset ($t-t_0$) or the closest value available. Errors are 1$\sigma$, while upper limits are 3$\sigma$. Whenever possible we specify the formal measured flux at the position of the afterglow for detection limits in brackets, when not available we have written 0.00. For observations performed more than 6 months from the burst, obtained to look for the host and not for the afterglow, we do not give the epoch and just mark them as ``Late''. 
\label{table:mmdata}  }      \\   
\hline\hline
GRB		& Redshift & Observatory	&  $t-t_0$		&	Band	&	Flux density 	&	Reference	\\
		&                 &			&  (days)		&	(GHz)	&	(mJy)		&		\\
\hline                        
\endfirsthead
\caption{continued.}\\
\hline\hline 
GRB		& Redshift & Observatory	&  $t-t_0$		&	Band	&	Flux density 	&	Reference	\\
		&                 &			&  (days)		&	(GHz)	&	(mJy)		&		\\
\hline                        
\endhead
\hline
\endfoot
970111  & ---  &BIMA       & 30.180 & 85.0 & $< 30.00$ ($   0.00 \pm 10.00$) &\citealt{smi97}                \\
        &      &BIMA       & 33.190 & 85.0 & $<  4.20$ ($   0.00 \pm  1.40$) &\citealt{smi97}                \\
\noalign{\smallskip}
\hline
\noalign{\smallskip}
970228  &   0.69 &BIMA       &  5.970 & 85.0 & $< 16.50$ ($   0.00 \pm  5.50$) &\citealt{smi97}                \\
        &      &BIMA       &  6.990 & 86.4 & $<  3.60$ ($   0.00 \pm  1.20$) &\citealt{smi97}                \\
\noalign{\smallskip}
        &      &JCMT       &Late    &350.0 & $<  3.09$ ($  -0.76 \pm  1.03$) &\citealt{ber03}                \\
        &      &JCMT       &Late    &350.0 & $<  3.96$ ($   1.78 \pm  1.32$) &\citealt{tan04}                \\
\noalign{\smallskip}
        &      &JCMT       &Late    &670.0 & $< 54.90$ ($ -21.40 \pm 18.30$) &\citealt{ber03}                \\
\noalign{\smallskip}
\hline
\noalign{\smallskip}
970508  &   0.83 &BIMA       &  1.296 & 85.0 & $< 82.80$ ($   0.00 \pm 27.60$) &\citealt{gru98}                \\
\end{longtable}
\begin{flushleft}
\tablefoottext{a}{The detection of GRB\,990123 at 353 GHz is probably due to a statistical fluctuation \citep{gal99}.}\\ 
\tablefoottext{b}{The detections of GRB\,010222 at 250 and 350 GHz are due to the host galaxy \citep{fra02}.}\\
\tablefoottext{c}{The detection of GRB\,050408 at 232 GHz is probably due to a statistical fluctuation \citep{deu07}.}\\
\end{flushleft}
}

\end{appendix}

\end{document}